\documentclass[journal,10pt,doublecolumn]{IEEEtran}
\usepackage{amsmath,amssymb,amsfonts,xpatch}
\usepackage{cite}
\usepackage[ruled,vlined]{algorithm2e}
\usepackage{setspace}
\usepackage{tikz}
\usepackage{graphicx}
\usepackage{cool}
\usepackage{caption}
\usepackage{graphics}
\usepackage{caption}
\usepackage{amsmath,amssymb,amsfonts,xpatch}
\usepackage{amsthm}
\usepackage{nccmath}
\usepackage{acronym}
\makeatletter
\let\NAT@parse\undefined
\makeatother
\usepackage[bookmarks,colorlinks,citecolor=green]{hyperref}

\acrodef{mac}[MAC]{multiple access channel} 
\acrodef{oimac}[OIMAC]{optical intensity multiple access channel} 
\acrodef{snr}[SNR]{signal-to-noise ratio}
\acrodef{awgn}[AWGN]{additive white Gaussian noise}  
\acrodef{mine}[MINE]{mutual information neural estimator}

\newtheorem{theorem}{Theorem}

\definecolor{blue}{rgb}{0,0,1}

\title{Neural Computation of Capacity Region of Memoryless Multiple Access Channels}
\author{\IEEEauthorblockN{Farhad~Mirkarimi,}
	\IEEEauthorblockA{\textit{Sharif University of Technology,}}
	\and
	\IEEEauthorblockN{and Nariman~Farsad,}
	\IEEEauthorblockA{\textit{Ryerson University}}
	\vspace{-0.5cm}
	
\thanks{This research is supported by Discovery Grant from the Natural Sciences and Engineering Research Council of Canada (NSERC) and Canada Foundation for Innovation (CFI), John R. Evans Leader Fund.}
\vspace{-0.2cm}
}
\begin{document}

\maketitle

\begin{abstract}
This paper provides a numerical framework for computing the achievable rate region of memoryless multiple-access channel (MAC) with a continuous alphabet from data. In particular, we use recent results on variational lower bounds on mutual information and KL-divergence to compute the boundaries of the rate region of MAC using a set of functions parameterized by neural networks. Our method relies on a variational lower bound on KL-divergence and an upper bound on KL-divergence based on the $f$-divergence inequalities. Unlike previous work, which computes an estimate on mutual information, which is neither a lower nor an upper bound, our method estimates a lower bound on mutual information. Our numerical results show that the proposed method provides tighter estimates compared to the MINE-based estimator at large SNRs while being computationally more efficient. Finally, we apply the proposed method to the optical intensity MAC and obtain a new achievable rate boundary tighter than prior works.  
\end{abstract}

\vspace{-0.2cm}
\section{introduction}
Computing closed-form expressions for the capacity region of general multi-user channels, except in few cases, has proved to be challenging. The capacity region of a two and three user memoryless \ac{mac} was formulated in \cite{ahl1973multi}. Later, \cite{msalehi} considered the capacity region of \ac{mac} with correlated sources and derived its achievable rates region.  In \cite{wolf} it was shown that feedback can increase the rate region of memoryless \ac{mac}, while \cite{permuter2009capacity} derived the capacity region of finite state \ac{mac}s and provided multi-letter expressions for capacity region. These prior works focused on Gaussian \ac{mac}s without fading or interference. In \cite{wyner} capacity region of K-user \ac{mac} over cellular systems was considered, where there is interference from adjacent cells, while \cite{tse1998} provides the optimal resource allocation scheme and the capacity region of Gaussian \ac{mac} with fading. 
Finally, a number of other works including \cite{oimac1, oimac2} consider the capacity of optical intensity \ac{mac} with both average and peak power constraints on inputs of the channel, and derive different inner and outer bounds for capacity region. Despite all these results, numerical computation of the capacity region of the arbitrary memoryless \ac{mac} has been challenging \cite{calvo10_BAMAC}.

Numerical computation of capacity most notably dates backs to the well-known work of Blahut and Arimoto \cite{1054855, 1054753}, where an alternating maximization method was used iteratively for computation of the capacity of the point-to-point discrete memoryless channel. There have been many follow-up works that extend this to other point-to-point channels. In \cite{voten}, an algorithm that maximizes mutual information for infinite-state indecomposable (noise-free) channel was considered for Markov sources, and in \cite{voten1}, the authors extend their work to the general finite-state channels. An extension of Blahut-Arimoto to channels with noncausal transmitter side information is provided in \cite{dup}, and \cite{caocapacity11} presented an algorithm called deterministic annealing for finding the capacity of discrete-time Poisson channel. The authors in \cite{perm} provide an extension of Blahut-Arimoto for estimating directed mutual information and use it for estimating the capacity of channels with feedback. Finally, in \cite{far20DAB} dynamic assignment Blahut-Arimoto (DAB) algorithm is introduced for efficiently evaluating the capacity of memoryless channels with continuous input and optimal input distribution with finite mass-points. 

Recently deep learning has become a new and powerful tool in communications with possible applications for designing new channel codes, new modulation schemes, and evaluating achievable information rates \cite{9085350}. Reinforcement learning is used to evaluate the capacity of point-to-point channels with feedback in \cite{aharonicomputing}. Recent advancements in estimating mutual information from data samples using deep learning methods \cite{belghazi2018mine} has resulted in several new works. In \cite{yechannel}, the estimator in \cite{belghazi2018mine} is used to produce efficient joint encoder and decoders for modulation (with a low probability of error) by maximizing mutual information between inputs and outputs of the channel. In \cite{aharoni20_CapMemChan}, a capacity estimation algorithm is developed for continuous channels with feedback using a deep learning-based estimator of directed information.

 Despite all these recent results, there has been little work on estimating the capacity region of multi-user channels.  In this work, we expand the 
  approach proposed in \cite{aharoni20_CapMemChan}, for estimating capacity of single user channels, to multi-user channels.  Moreover, we expand the neural estimators \cite{belghazi2018mine} and \cite{chan2019neural}, which were used in \cite{aharoni20_CapMemChan}, and provide a different lower bound estimator on mutual information. This is achieved by proposing different variational lower and upper bounds on KL-divergence and using these variational bounds to obtain a lower-bound on mutual information and conditional mutual information. We then describe a method using neural networks for estimating this bound from data, while maximizing it with respect to the input distribution. Our algorithm is computationally more efficient than the method proposed in \cite{aharoni20_CapMemChan}.  Numerical evaluations demonstrate that the estimation of this bound from data will be close to the optimal rates for the Gaussian \ac{mac}. We also show that our approach can be used to estimate a new and tighter bound for the optical intensity \ac{mac} \cite{oimac1, oimac2}. The proposed neural estimator for mutual information and conditional mutual information can also be used for design of end-to-end communication systems, including multi-user systems, and for representation learning in machine learning. 
 
 
\section{ Neural Estimation of Mutual information}
\label{sec:NeuralMIestimator}
In this section,  we present a method for computing lower bounds on mutual information using variational bounds. To do this, we first provide a lower and upper bound on KL-divergence in Section \ref{subsec:KLbounds}, and then use these bounds to present an estimator for the lower bound on mutual information in Section \ref{subsec:MIbounds}.

\vspace{-0.3cm}
\subsection{Bounds on KL-Divergence}
\label{subsec:KLbounds}
 First, we present a variational lower bound on KL-divergence using a similar approach to the one taken in \cite{2019variational} to lower bound mutual information. 

\begin{theorem}
	\label{thm:KLlower}
	Let $P \in \mathcal{P}(\mathcal{X})$ and $Q \in \mathcal{P}(\mathcal{X})$ be two probability measures on the random variables $X$ with the sample space $\mathcal{X}$, where $\mathcal{P}(\mathcal{X})$  denote the set of all probability measures over the Borel $\sigma$-algebra on $\mathcal{X}$. If $P$ is absolutely continuous with respect to $Q$, then the KL-divergence between the two distributions $P$ and $Q$ is lower bounded by: 
	
	\begin{equation}\label{imj}
			D(P || Q) \geq \mathbb{E}_{P}[T(x)] - \frac{\mathbb{E}_Q\left[e^{T(x)}\right]}{a} - \log(a) + 1,
	\end{equation}
	for all integrable functions $T(x)$ and $a \geq 0$.
\end{theorem}
\begin{IEEEproof}
	From \cite{donsker1983, belghazi2018mine, 2019variational} we have: 
	\begin{flalign}
		D(P||Q) \geq \mathbb{E}_{P}[T(x)]-\log\left(\mathbb{E}_{Q}[e^{T(x)}]\right),\label{nbvx}
	\end{flalign}
	where $T(x)$ is any function $T: \mathcal{X} \rightarrow \mathbb{R}$ that satisfies the integrability constraints of Theorem \ref{thm:KLlower}.  
	Now using the inequality $\log(u)\leq \frac{u}{\alpha}+\log(\alpha)-1$ with $\alpha\geq0 $ and $u=\mathbb{E}_{Q}[e^{T(x)}]$ in \eqref{nbvx}
	we conclude \eqref{imj}.
\end{IEEEproof}
%


Note that the equality in the lower bound is tight when $T(x)= \log (\frac{dP}{dQ}) + c$, for some constant $c$. To find the $a$ and $T(x)$ that  maximize this lower bound on KL-divergence, one can represent $T(x)$ by neural networks $\phi_{\theta_T}(x)$ with parameters $\theta_T$, as a trainable parameter .  Using the variational methods, and a loss function that will be described in the next section, we can maximize the lower bound on KL-divergence in \eqref{imj}. By using enough data and a large batch size, it is theoretically possible to find a tight lower bound on KL-divergence between $P$ and $Q$ as data size goes to infinity. Note that it is possible to extend Theorem \ref{thm:KLlower} to the case where $P$ and $Q$ are joint distributions of multiple random variables. 



For the upper bound on KL-divergence, we use the following inequality presented in
\cite[Theorem 20]{verdu2016fDiverg}, which is restated here for convenience.
\begin{theorem}
	\label{thm:KLupper}
	Let $P$ and $Q$ be two distributions defined in Theorem \ref{thm:KLlower}. Then we have the following upper bound on the KL-divergence: 

\begin{align}
    \label{upper1}
		D(P||Q)  \leq & \log(1 + \chi^{2}(P||Q)) \nonumber \\ 
		& \qquad -\frac{1.5(\chi^{2}(P||Q))^{2} \log e}{(1 + \chi^{2}(Q||P))(1 + \chi^{2}(P||Q))^2 - 1} \nonumber \\
		\triangleq & \chi^{2}_{\mathsf{UP}}(P||Q),
\end{align}
where $\chi^{2}(P||Q)$ represents $\chi^{2}$ distance between $P$ and $Q$ distributions defined as:
\begin{equation}
\chi^{2}(P||Q)=\int_{}^{}\frac{(p(x)-q(x))^{2}}{q(x)}dx
\end{equation}

\end{theorem}
\begin{IEEEproof}
	See \cite[Theorem 20]{verdu2016fDiverg} for the proof.
\end{IEEEproof}
%

In the next section, we use these two bounds to derive a variational lower bound on mutual information. 


\vspace{-0.5cm}
\subsection{Estimation of Entropy and Mutual Information}
\label{subsec:MIbounds}

We begin the section by presenting variational bounds on entropy.   To estimate these bounds using the variational methods  \cite{chan2019neural}, a reference (and arbitrary) distribution $Q$ over the random variable $X$ with pdf $q(x)$ is used in place of the true and unkown distribution $P$ with pdf $p(x)$. Using $Q$, the entropy of the random variable $X$ can be written as:
\begin{equation}\label{mes}
	h(X)=\mathbb{E}_P [-\log(q(x))]-D(P||Q).
\end{equation}
Note that the first term is the cross-entropy term $h_{\text{CE}}(P,Q)$. Now we seek to find a variational bound on this expression. The second term, which is the KL-divergence, could be bounded using Theorem \ref{thm:KLlower} that was provided in the previous section, which results in an upper bound on entropy of $X$ given by:

{\small
	\vspace{-0.6cm}
	\begin{flalign*}
		h(X) \leq &\mathbb{E}_P [- \mspace{-3mu}\log(q(x))]  \mspace{-3mu}- \mspace{-3mu}  \mathbb{E}_{P}[T(x)]  \mspace{-3mu}+ \mspace{-3mu} \frac{\mathbb{E}_Q\left[e^{T(x)}\right]}{a}  \mspace{-3mu} +  \mspace{-3mu} \log(a)  \mspace{-3mu}- \mspace{-3mu} 1.  
	\end{flalign*}
}%
To make the upper bound tight, a neural network $\phi_{\theta_T}(x)$ is used to represent $T(x)$, and the parameters $\theta_T$ are optimized using the loss function
\begin{equation}\label{eq:loss}
	L(\theta)=-\mathbb{E}_P[\phi_{\theta_T}(x)]+ \frac{\mathbb{E}_Q[e^{\phi_{\theta_T}(x^\prime)}]}{\alpha_a} + \log(\alpha_a) - 1,
\end{equation}
where $x \sim p(x)$ and $x^\prime \sim q(x^\prime )$.  Note that the expectation terms are evaluated using empirical averaging over the samples in the~minibatch and the parameter $\alpha_a$ can be tuned as a hyperparameter or be treated as a trainable parameter. 


We now present a method to compute the variational lower bound on mutual information. Using \eqref{mes}, and i.i.d. reference random variables $X^{'}$ and $Z^{'}$, mutual information is represented as

{\small
	\vspace{-0.2cm}
	\begin{align}
		I(X;Z) &= h(X) + h(Z) - h(X,Z), \nonumber \\
		& =D(P_{X,Z}||Q_{X^{'},Z^{'}})-D(P_X||Q_{X^{'}})-D(P_Z||Q_{Z^{'}}). \label{eq:MIasKLs}
	\end{align}
}%
Note that since we can choose the reference distributions to be i.i.d., the cross-entropy terms in \eqref{mes} will cancel out leaving only the KL-divergence terms. In addition \cite{aharoni20_CapMemChan} computes \eqref{eq:MIasKLs} by using  Donsker-Varadhan bound for all terms (even the negative terms), which results in neither a lower or an upper bound.  For finding variational lower bound on $I(X;Z)$, we lower bound the first term in \eqref{eq:MIasKLs} using Theorem \ref{thm:KLlower} and upper bound the next two terms using Theorem \ref{thm:KLupper}. Therefore, we have 

{\small
\vspace{-0.2cm}
\begin{align}
       I(X;Z)\geq &
		\underbrace{\mathbb{E}_{P_{X,Z}}\mspace{-3mu}[T(X,Z)] \mspace{-3mu}-\mspace{-3mu} \frac{\mathbb{E}_{Q_{X^{'},Z^{'}}}\mspace{-3mu}\left[e^{T(X^{'},Z^{'})}\mspace{-3mu}\right]}{a(z^{'})}\mspace{-3mu}-\mspace{-3mu}\log(a(z^{'}))\mspace{-3mu}+\mspace{-3mu}1}_\text{Variational lower bound on $D(P_{X,Z}||Q_{X^{'},Z^{'}})$} \nonumber \\
		& -\underbrace{\chi^{2}_{\mathsf{UP}}(P_{X}||Q_{X^{'}})}_\text{$\chi^{2}$  upper bound on $D(P_{X}||Q_{X^{'}})$}
		-\underbrace{\chi^{2}_{\mathsf{UP}}(P_{Z}||Q_{Z^{'}})}_\text{$\chi^{2}$ upper bound on $D(P_{Z}||Q_{Z^{'}})$}. \label{eq:MILB}
\end{align}
}%

Note that $Z^{'}$ and $X^{'}$ are reference random variables, and in this work they are uniformly distributed over the support of $X$ and $Z$ in each batch as was suggested in \cite{aharoni20_CapMemChan, chan2019neural}. We estimate the $\chi^{2}$ distance terms in \eqref{eq:MILB} using a histogram-based density estimation as follows. Assume that we have two histogram estimates of probability  density functions $P$ and $Q$ with $m$ bins. Let the estimated probability corresponding to bin $i$ for $P$ and $Q$ be $f(i)$ and $g(i)$, respectively. In this case we have:
\begin{equation}
\chi^{2}(P||Q)=\sum_{i=1}^{m}\frac{(f(i)- g(i))^{2}}{g(i)}.
\end{equation}
This method is also applied for computation of $\chi^{2}$-divergence between joint distributions. To maximize the lower bound and make it tighter, neural networks are used to represent the function $T$ in \eqref{eq:MILB}, which is then trained to minimize the negative of the loss function in \eqref{eq:loss} (i.e., maximize the loss in \eqref{eq:loss}) by setting $x$ and  $x^\prime$ to $(x,z)$ and $(x^\prime,z^\prime)$, respectively.  The expectations in \eqref{eq:MILB} are estimated using the empirical average over the minibatch.  


%



\vspace{-0.1cm}
\section{Estimating Achievable Region of MAC}
\vspace{-0.1cm}
\label{sec:NNmacCap}
In this section, we provide a numerical framework for computing inner bounds on the capacity region of memoryless \ac{mac}. To simplify the presentation we focus on a two-user \ac{mac}, but the framework can be extended to larger number of users.  In a 2-user memoryless \ac{mac}, if we denote the input distribution of the first, and second user by $X$ and $Y$, and the output distribution by $Z$, then each user's rate, $R_1$ and $R_2$ respectively, satisfies the following conditions:

{\small	
\begin{flalign*}
R_1\leq I(X;Z\mid Y), ~~ R_2\leq I(Y;Z\mid X), ~~
R_1+R_2\leq I(X,Y;Z).
\end{flalign*}
}%
We continue the rest of this section by describing our {\em neural achievable rate region (NARR)} estimator that estimates tight lower bounds on $I(X;Z|Y)$, $I(Y;Z|X)$, and $I(X,Y;Z)$. Then we describe how we optimize the input distributions to maximize boundaries of the achievable rate regions in an iterative manner. 

\vspace{-0.1cm}
\subsection{NARR Estimator}
\vspace{-0.1cm}
In order to compute boundaries of the capacity region in the two-user \ac{mac}, we need to estimate $I(X;Z|Y)$, $I(Y;Z|X)$, and $I(X,Y;Z)$. First, we focus on $I(X;Z|Y)$ for presenting our algorithm, which can also be applied to $I(Y;Z|X)$.  

We begin by expanding the conditional mutual information in terms of a number entropy terms. Specifically, we have
\begin{align}
I(X;Z\mid Y)= & h(Y,Z)+h(X,Y)-h(Y)-h(Y,X,Z) \nonumber \\
=&D(P_{X,Y,Z}||Q_{X^{'},Y^{'},Z^{'}})+D(P_{Y}||Q_{Y^{'}}) \nonumber \\
&\quad -D(P_{X,Y}||Q_{X^{'},Y^{'}})-D(P_{Y,Z}||Q_{Y^{'},Z^{'}}), \label{eq:condMIKL}
\end{align}
where that last equality follows from \eqref{mes}, using i.i.d. reference variable $X^{'}$, $Y^{'}$, and $Z^{'}$, and canceling out the cross-entropy terms. 
Applying Theorem \ref{thm:KLlower} to the first two terms of \eqref{eq:condMIKL} and Theorem \ref{thm:KLupper} to the last two terms results in 
{\small
\begin{flalign}
I(X;Z|Y) \geq & \mathbb{E}[T_{\theta_{T1}}^{(1)}\mspace{-3mu} (x,y,z)] \mspace{-3mu} - \mspace{-3mu} \frac{\mathbb{E} \mspace{-3mu} \left[e^{T_{\theta_{T1}}^{(1)}\mspace{-3mu}(x^\prime\mspace{-3mu}, y^{\prime} \mspace{-3mu}, z^{\prime})}\right]}{\alpha^{(1)}} \mspace{-3mu}
- \mspace{-3mu} \log(\alpha^{(1)}) \mspace{-3mu} + \mspace{-3mu} 1 \nonumber \\ 
 &\quad +\mathbb{E}[T_{\theta_{T2}}^{(2)}(y)] - \frac{\mathbb{E}\left[e^{T_{\theta_{T2}}^{(2)}(y^{\prime})}\right]}{\alpha^{(2)}} - \log(\alpha^{(2)}) + 1 \nonumber \\
 & \quad - \chi^{2}_{\mathsf{UP}}(P_{X,Y}||Q_{X^{'},Y^{'}}) - \chi^{2}_{\mathsf{UP}}(P_{Y,Z}||Q_{Y^{'},Z^{'}}) 
\label{eq:condMIvarLB}
\end{flalign}
}%
where $T_{\theta_{T1}}^{(1)}$ and $T_{\theta_{T2}}^{(1)}$ are two functions parameterized by two distinct neural networks. Here for brevity we have omitted from the notation the distribution over which the expectations are taken. The reader may revisit the formulations in the previous section for this information. The variational upper bound terms in \eqref{eq:condMIvarLB} can be evaluated once from the training data as described in the previous section. The expectation terms can all be estimated using empirical averages over a minibatch of size $N$. The loss function used to maximize the lower bound on the conditional mutual information is 

{\small
\begin{flalign}
	&\mathcal{L}_1(\theta_{T1}, \theta_{T2}, \alpha^{(1)}, \alpha^{(2)}) =L_{1}(\theta_{T1}, \alpha^{(1)})+L_{2}(\theta_{T2}, \alpha^{(2)}), \nonumber \\
&L_{1}(\theta_{T1}, \alpha^{(1)}) = - \frac{1}{N}\sum_{i=1}^{N}T_{\theta_{T1}}^{(1)}(x_i,y_i,z_i)  \nonumber \\ 
& \qquad \qquad  \qquad  \qquad + \frac{\frac{1}{N}\sum_{i=1}^{N}T_{\theta_{T1}}^{(1)}(x^{'}_i,y^{'}_i,z^{'}_i)}{\alpha^{(1)}} +\log(\alpha^{(1)}), \nonumber \\
&L_{2}(\theta_{T2}, \alpha^{(2)}) \mspace{-3mu} =  \mspace{-3mu} - \mspace{-3mu} \frac{1}{N}\sum_{i=1}^{N}T_{\theta_{T2}}^{(2)}(y_i) \mspace{-3mu} + \mspace{-3mu} \frac{\frac{1}{N}\sum_{i=1}^{N}T_{\theta_{T2}}^{(2)}(y^{'}_i)}{\alpha^{(2)}} \mspace{-3mu}+\mspace{-3mu}\log(\alpha^{(2)}). \label{eq:lossCondiMI}
\end{flalign}
}%
Once the parameters are trained, the lower bound on conditional mutual information can be estimated from $M$ samples using \eqref{eq:condMIvarLB}, where $M$ can be much larger than $N$. 

To estimate the conditional mutual information $I(Y;Z|X)$ the same approach can be used with a separate set of neural networks $S_{\theta_{S1}}^{(1)}$ and $S_{\theta_{S2}}^{(2)}$. The loss function for $I(Y;Z|X)$ is then given by $\mathcal{L}_2(\theta_{S1}, \theta_{S2}, \beta^{(1)}, \beta^{(2)})$, which are defined similar to \eqref{eq:lossCondiMI}.


Following the same approach we can obtain a variational bound on $I(X,Y;Z)$ given by
%

	\begin{flalign}
		I(X,Y; Z)  \geq & \mathbb{E}[U_{\theta_{U}}\mspace{-3mu} (x,y,z)] \mspace{-3mu} - \mspace{-3mu} \frac{\mathbb{E} \mspace{-3mu} \left[e^{U_{\theta_{U}}\mspace{-3mu}(x^\prime\mspace{-3mu}, y^{\prime} \mspace{-3mu}, z^{\prime})}\right]}{\gamma} \mspace{-3mu} - \mspace{-3mu} \log(\gamma) \mspace{-3mu} + \mspace{-3mu} 1\nonumber \\
		& \qquad - \chi^{2}_{\mathsf{UP}}(P_{X,Y}||Q_{X^{'},Y^{'}}) - \chi^{2}_{\mathsf{UP}}(P_{Z}||Q_{Z^{'}}) 
\label{eq:multiMIvarLB}
	\end{flalign}
where $U_{\theta_{U}}$ is a neural network with parameters $\theta_{U}$, is trainable variable. We can employ the same technique applied to the conditional mutual information to define a loss $\mathcal{L}_3(\theta_{U}, \gamma)$ and using training to maximize the lower bound with respect to  $\theta_{U}$.

To estimate an inner bound on the capacity of the \ac{mac}, NARR estimates $I(X;Z|Y)$, $I(Y;Z|X)$, and $I(X,Y;Z)$ jointly by using the loss function $\mathcal{L} = \mathcal{L}_1+\mathcal{L}_2+\mathcal{L}_3$.


\begin{algorithm}[t]
	\caption{Neural MAC inner bound estimator}\label{alg}
	\KwIn{Channel model or its GAN approximation}
	\KwOut{Estimate of the inner capacity region of MAC}
	Initialize parameters of NARR and NIT randomly \\
	\While{not converged or max iteration not reached}
	{
		\textbf{Phase 1}: \textbf{Train NARR} \\
		Generate $B$ sample of $N_1, N_2$: $\{(n_1^{(i)},n_2^{(i)})\}_{i=1}^B$ \\ 
		Generate $\{(x_i, y_i, z_i)\}_{i=1}^B$ using NIT and channel \\
		Calculate the NARR loss $\mathcal{L}$ for this batch \\
	
	    Train $S, T, U$ using gradient-descent \\
		\textbf{Phase 2}: \textbf{Train NIT}\\
		Generate $B$ sample of $N_1, N_2$: $\{(n_1^{(i)},n_2^{(i)})\}_{i=1}^B$ \\ 
		Generate $\{(x_i, y_i, z_i)\}_{i=1}^B$ using NIT and channel \\
		Use channel and NARR to calculate:
		 $$\hat{\mathcal{I}} = \hat{I}(X;Z\mid Y) + \hat{I}(Y;Z\mid X) + \hat{I}(X,Y;Z)$$ \\
		Use  $-\hat{\mathcal{I}}$ as loss to train NIT
	}
	Perform final evaluation on all  or subset of data \\
	\textbf{Return}: $\hat{I}(X;Z\mid Y), \hat{I}(Y;Z\mid X), \hat{I}(X,Y;Z)$	
\end{algorithm}

\subsection{Estimating the Optimal Input Distributions}
\label{sec:MaxInputDist}
In the previous section we presented NARR, which is a data-driven approach for estimating various information theoretic quantities that appear in capacity region of a two-user \ac{mac}, for a specific input distribution. This data-driven approach can yield tight lower bounds on corresponding mutual information terms if a proper network architecture and large number of samples and batch sizes are used for training. However, to find an inner capacity region, these information theoretic quantities must be maximized with respect to the channel input distributions.

Inspired by generative networks (GAN) \cite{GAN} and \cite{aharoni20_CapMemChan}, we use a separate generative neural network that uses two i.i.d. random variables from a known distribution ($N_1$ and $N_2$) as seeds, and outputs the channel inputs $X$ and $Y$. We call this network the {\em neural input transformer (NIT)}. Note that the NIT can approximate a wide range of channel input distributions. This is similar to the generative neural network, called normalizing flow, which transforms a known distribution to another by training the generator such that its output moves towards the target distribution. 

The NARR and NIT are trained iteratively to find a lower-bound on achievable rate region of memoryless \ac{mac}. A single training iteration has 2 phases. In phase 1, the weights of the NIT network are kept constant and the NARR network is trained. In our experiments we have found that setting $\alpha$, $\beta$, $\gamma$ to an appropriate constant value as explained in numerical section gives the best performance..  Note that the same batch is used in training phase 1.1 and 1.2.  In phase 2, a new batch is generated and the NARR network is kept constant while the NIT network is trained. The training continues in this fashion until the estimates $I(X;Z|Y)$, $I(Y;Z|X)$, and $I(X,Y;Z)$ converge or until a specific number of iterations are reached. Algorithm \ref{alg} summarizes this training procedure. 


\section{Numerical Results}
This section evaluates the proposed method on two different \ac{mac}s. The implementations are available at GitHub repository of paper\footnote{\url{https://github.com/Farhad-Mrkm/Neural-Capacity-Computation}}. First, the \ac{awgn} \ac{mac} is considered and it is shown that the proposed approach is a better estimator of the optimal user rates and the sum rate compared to the method based on \ac{mine} \cite{belghazi2018mine} in high SNR regimes. Note that the capacity region of the \ac{awgn} \ac{mac} is known and our goal in this first experiment is to evaluate the performance of the proposed approach. Second, the \ac{oimac} is considered. For this channel the capacity region is unknown. Hence, we compare our inner bound estimator to recent inner and outer bounds derived in \cite{oimac1} and \cite{oimac2}, and show that using our approach a tighter bound can be estimated compared to these prior work. 

For all the experiments, each of the neural networks in NARR is a four-layer feedforward (FF) neural network with 64 hidden dimensions and the ReLU activation function. 
Each reference random variable (e.g., $X^{'}$) is a uniform random variable over the support of the corresponding variable it is representing (e.g., $X$). The support is taken to be the maximum and minimum of the corresponding variable in the minibatch as suggested in \cite{chan2019neural}. In the \ac{mine}-based approach, we use a similar technique proposed in \cite{aharoni20_CapMemChan}, where all the KL-divergence terms in $I(X;Z|Y)$, $I(Y;Z|X)$, and $I(X,Y;Z)$ (e.g., see \eqref{eq:condMIKL}) are estimated using \ac{mine} \cite{belghazi2018mine}. 
The NIT consists of six-layer FF neural network. The input to the NIT is $[N_1, N_2]$ where $N_1$ and $N_2$ are i.i.d. standard Gaussians. First four-layer uses ReLU activation function. The last two layers are normalization layers to enforce the average and peak power constraints. 
Outputs of the NIT ($X$,$Y$) is fed to the channel to produce the channel output $Z$.

\begin{figure}
	\includegraphics[width=3in]{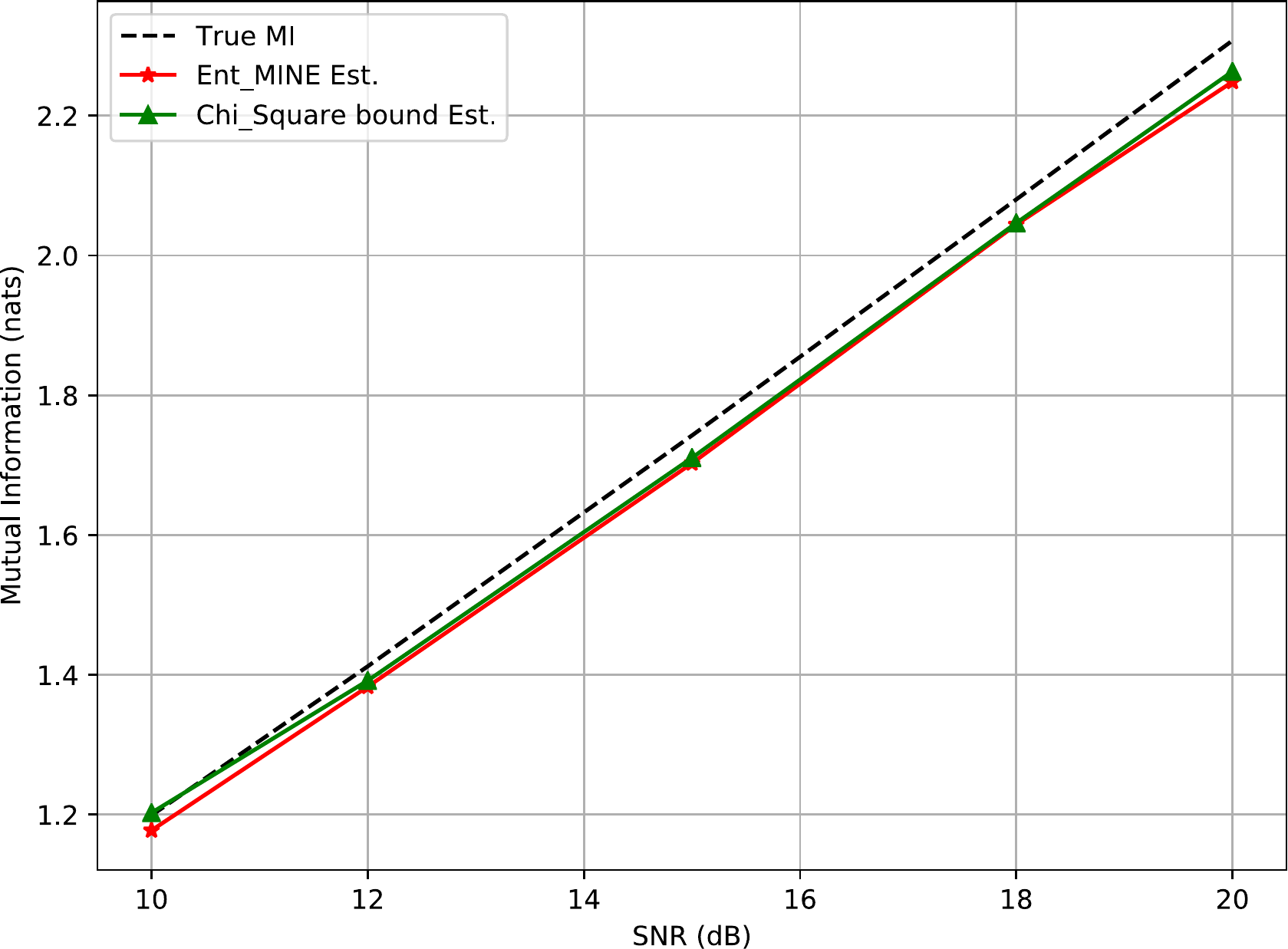}
	\caption{The optimal achievable rates of the first user $I(Y;Z\mid X)$ \ac{awgn} \ac{mac}.}
	\label{fig:I_YZXplot}
	\vspace{-0.2cm}
\end{figure}

\begin{figure}
	\includegraphics[width=3in]{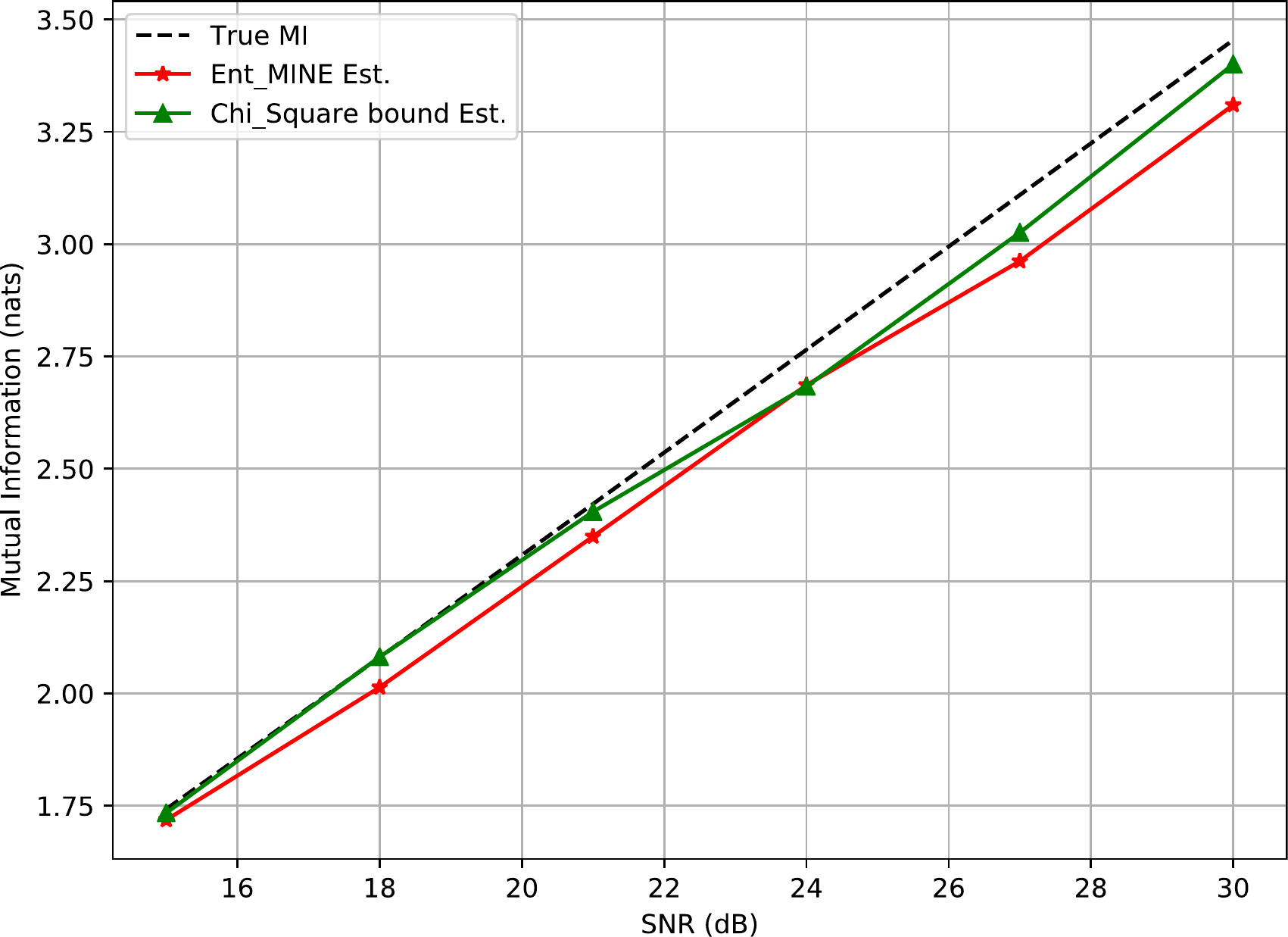}
	\caption{The optimal achievable rates of the second user $I(X;Z\mid Y)$ \ac{awgn} \ac{mac}.}
	\label{fig:I_XZYplot}
	\vspace{-0.2cm}
\end{figure}

\begin{figure}
	\includegraphics[width=3in]{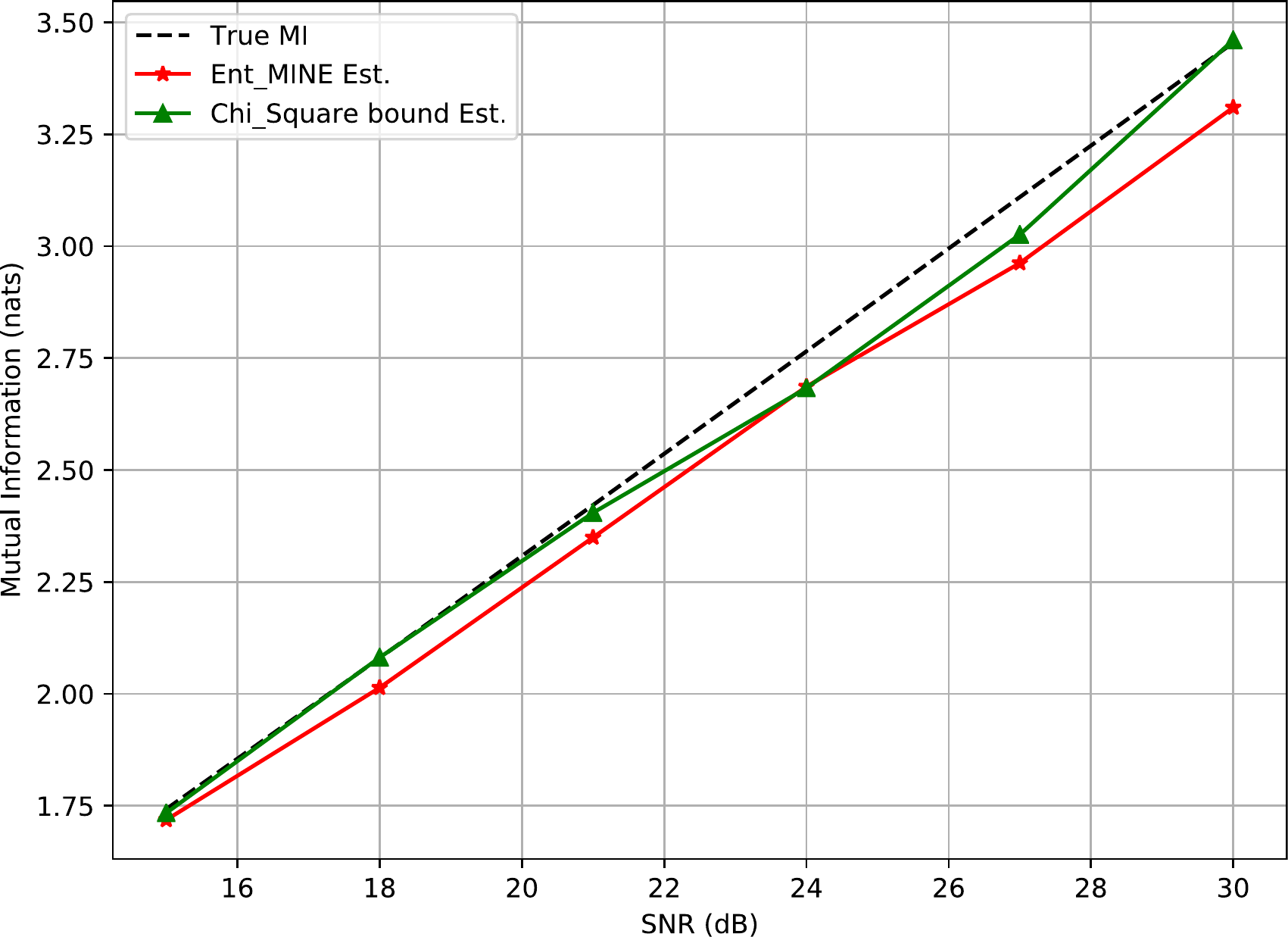}
	\caption{The optimal sum rate $I(X,Y;Z)$ for \ac{awgn} \ac{mac}.}
	\label{fig:I_XYZplot}
	\vspace{-0.2cm}
\end{figure}

For optimization we use the Adam optimizer with a learning rate $10^{-5}$ to $10^{-3}$. The mini-batch size is chosen 9,000 for SNRs above 20dB and 12,000 otherwise. In our experiments, we have found that it suffices to choose the parameters $\alpha$, $\beta$, and $\gamma$ in the range (0.9,3) for the algorithm to converge quickly. Here we use the value of 2. Here, we treat this parameter as a hyperparameter and tune it for each \ac{snr}. In general, training using the proposed approach takes less time compared to the \ac{mine}-based approach where two more neural networks must be optimized for each of $I(X;Z|Y)$, $I(Y;Z|X)$, and $I(X,Y;Z)$. The reason is that computation of upper bounds are done outside training loop. In a single GPU training our bounds for each SNR point takes around 5 minutes. 
In general, using neural estimation does not require days or hours of training, and could produce good estimates for the channels considered in minutes.

We now present the results. First, we consider the \ac{awgn} \ac{mac} where users 1 and 2 have the power constraints $P_{1}=E[X^{2}]\leq 30$dB, and $P_{2}=E[Y^{2}]\leq 20$dB, respectively.  To train our estimator and obtain the estimates, we use $6000$ samples and draw the batches from these samples. Also at highest \acp{snr} we use the learning rate of $10^{-3}$ and decrease the learning rate to $10^{-5}$ in the highest \ac{snr}. For computation of different \ac{snr} point we fix noise variance to one and increase power respectively. 
The achievable rates of each user and sum capacity of \ac{mac} are shown in Figs. \ref{fig:I_YZXplot}-\ref{fig:I_XYZplot}. As can be seen, the proposed approach results in tighter estimated bounds compared to the \ac{mine}-based approach, while maintaining lower complexity. This is especially evident at higher \acp{snr}. 

\begin{figure}
	\includegraphics[width=3.4in]{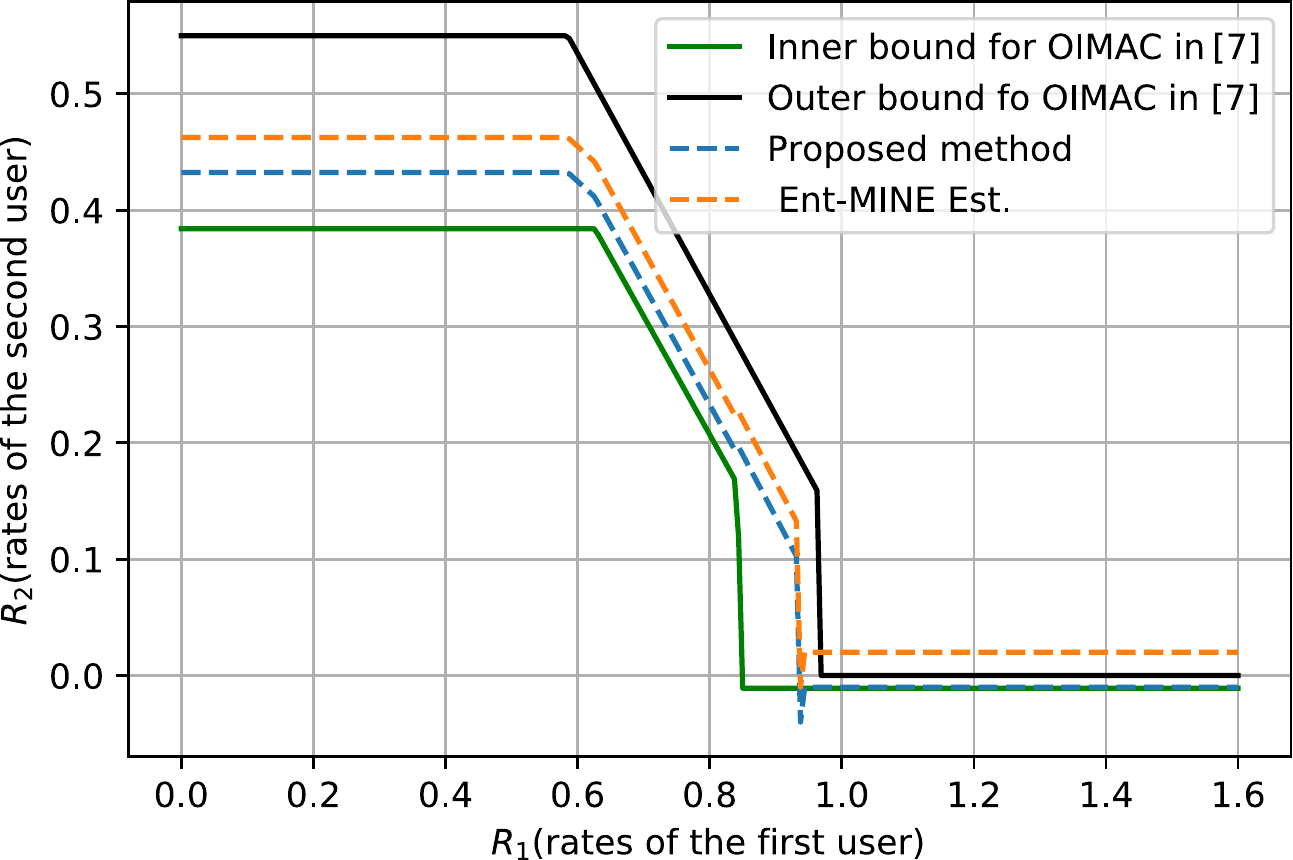}
	\caption{ The outer and inner bounds of \ac{oimac} from \cite{oimac1} compared to our proposed neural lower-bound estimator and \ac{mine}-based estimator. In the numerical evaluations, it is assumed that $\frac{\mathbb{E}[X]}{A_2}=\frac{\mathbb{E}[Y]}{A_1}=.2$, $A_1=10$dB , $A_2=5$dB, and $\sigma^2=1$}
	\label{fig:ratesIOMAC}
\end{figure}


Second, we consider \ac{oimac}, which is defined as:
\begin{flalign*}
	&Z = X + Y + N, \quad  N\sim \mathcal{N}(0,\sigma^{2}),\quad 0\leq Y\leq A_{1},\\
	&P_1 = \mathbb{E}[Y^{2}]\leq \epsilon_{1},\quad P_2 = \mathbb{E}[X^{2}]\leq \epsilon_{2},\quad 0\leq X\leq A_{2}
\end{flalign*}
Since the capacity region of \ac{oimac} is unknown, we rely on recent inner and outer bounds derived in \cite{oimac1}, \cite{oimac2} for comparison. Fig.~\ref{fig:ratesIOMAC} depicts the achievable inner bound obtained using our approach and the bound obtained using \ac{mine} and compares them with the inner and the outer bounds of \cite{oimac1}. The neural estimations achieve a better inner bound compare to the bound presented in \cite{oimac1}. However, we observe that at $R_1 \geq 0.9$, the \ac{mine}-based estimator goes over the upper bound. This can be due to numerical instabilities or because \ac{mine}-based estimator is neither an upper bound nor a lower bound on capacity. Regardless, both neural estimators show that the as $R_1$ rate goes above 0.9, the $R_2$ rate drops towards zero. This demonstrates that the neural estimators can be used to find new and better bounds for channels where the capacity region is not known. But one must be aware of numerical instabilities that might exist at extremely small or large rates.


\section{Conclusions}
We proposed a new approach for evaluating a lower bound on mutual information and conditional mutual information using neural networks and the variational methods. It was shown that this technique can be used to estimate the inner bound on the capacity region of the \ac{mac} directly from data, by using the NIT network that learns the channel input distribution that maximizes the inner bound. The estimated bounds were shown to be tighter than prior methods based on \ac{mine}. Moreover, the proposed method exhibits lower computational complexity compared to the \ac{mine}-based approach. We must caution that these results while promising are still preliminary and more investigation is needed to establish that neural network based estimation of bounds on the capacity region of \ac{mac} can be reliable and stable at all rates.  As part of future work, we will further improve our upper bound in KL-divergence and will also explore new bounds that can reduce sample complexity and result in better estimators with theoretical guarantees. We will also explore if the proposed method can be employed to learn new channel codes for the \ac{mac}.

\bibliographystyle{IEEEtran}
\bibliography{mains1}

\end{document}